\documentstyle[preprint,revtex]{aps}

\tightenlines

\begin{document}

\preprint{SFU HEP-101-93}
\preprint{\ }
\preprint{TRIUMF 93-15}
\preprint{\ }
\preprint{hep-lat/yymmnn}
\preprint{\ }
\preprint{March 1993}

\draft

\begin{title}
Flux-tubes in three-dimensional lattice gauge theories
\end{title}

\author{Howard D. Trottier}
\begin{instit}
Department of Physics, Simon Fraser University,
Burnaby, B.C., Canada V5A 1S6\thinspace\cite{HDT} \\
and \\
TRIUMF, 4004 Wesbrook Mall, Vancouver, B.C., Canada V6T 2A3
\end{instit}

\author{R. M. Woloshyn}
\begin{instit}
TRIUMF, 4004 Wesbrook Mall, Vancouver, B.C., Canada V6T 2A3
\end{instit}

\begin{abstract}
Measurements of flux-tubes generated by sources
in different representations of SU(2) and U(1)
lattice gauge theory in three dimensions are reported.
Heavy ``quarks'' are considered
in three representations of SU(2):
fundamental ($j=1/2$), adjoint ($j=1$), and quartet ($j=3/2$).
Wilson loops are used to introduce a static
quark-antiquark ($Q_j \overline Q_j$) pair. Several
attributes of the fields generated by the $Q_j \overline Q_j$
pair are measured. In particular,
the first direct lattice measurements of the flux-tube
cross-section ${\cal A}_j$ as a function of representation
are made. It is found that ${\cal A}_j \approx {\rm constant}$,
to within about 10\% (a rough estimate of the overall quality
of the data). The results are consistent
with a connection between the string tension $\sigma_j$
and cross-section suggested by a simplified
model of flux-tube formation,
$\sigma_j = g^2 j(j+1) / (2 {\cal A}_j)$
[where $g$ is the gauge coupling],
given that the string tension scales like the
Casimir $j(j+1)$, as observed in previous lattice
studies in both three and four dimensions
(and confirmed here up to the quartet representation).
These results can be used discriminate among phenomenological
models of the physics underlying confinement.
Flux-tube measurements are also made in
compact QED$_3$, which exhibits electric
confinement due to magnetic monopole condensation.
Singly- and doubly-charged Wilson loops
are considered. The string tension is found to scale
like the squared-charge, and the flux-tube cross-section
is found is be independent of the charge, to a good
approximation. The results of these three-dimensional
SU(2) and U(1) simulations taken together lend some support,
albeit indirectly, to a conjecture that the
dual superconductor mechanism underlies confinement
in compact gauge theories in both three and four dimensions.
\end{abstract}

%%% This command resulted in
%%% a blank page with only "Typset in REVTEX"
%%% \pacs{1990 PACS number(s): }
\newpage

\narrowtext

\section{\bf Introduction}

Flux-tube formation provides an attractive description
of confinement in quantum chromodynamics (QCD).
In a simplified flux-tube picture of a very heavy quark-antiquark
pair, color-electric field lines
running between the quarks are assumed to be ``squeezed''
into a cylinder whose cross-section is independent
of the quark separation $R$ (ignoring end effects),
resulting in linear confinement.

Consider heavy ``quarks'' in an arbitrary representation of
the gauge group, which we take to be SU(2) for convenience.
The color-electric field $E_j$ between quarks
in the $j$-th representation is determined, in an Abelian
approximation, by Gauss' Law \cite{HasKuti}
$E_j {\cal A}_j = g Q_j$, where the quark
color ``charge'' $Q_j$ is related to the group Casimir,
$Q_j^2 = j(j+1)$, $g$ is the gauge coupling,
and ${\cal A}_j$ is the cross-section of the flux tube.
In this simplified model, $E_j$ is assumed to be constant
across the flux-tube cross-section. In the case of non-fundamental
representation sources, $R$ is also assumed to be below the
threshold for fission of the tube.
The interaction energy of the system
$V_j^{\rm int}(R) = \case{1}/{2} E_j^2 {\cal A}_j R = \sigma_j R$,
and thus the string tension $\sigma_j$ is given by
\begin{equation}
   \sigma_j = { g^2 j(j+1) \over 2 {\cal A}_j } .
\label{tube}
\end{equation}
In more a detailed flux-tube model, the color-field $E_j$ may
vary in magnitude across the flux-tube cross-section;
the cross-section in Eq. (\ref{tube}) is then defined by
${\cal A}_j \equiv ( \int E_j dA_j )^2 / \int E_j^2 dA_j$;
it is also possible to relax the constraint that
${\cal A}_j$ is independent of $R$.

If the general features of the flux-tube model of confinement
are consistent with QCD, then the connection Eq. (\ref{tube})
between $\sigma_j$ and ${\cal A}_j$ should hold.
However, within the context of the flux-tube picture,
${\cal A}_j$ is an unknown function of the quark representation.
The cross-section is determined by the fundamental
dynamics of QCD. Flux-tube formation has been observed in
lattice QCD simulations in four dimensions
\cite{Flux1,Flux2,Sommer,HayWos,HayWosFt,Cooling}
and as well as in three dimensions \cite{Caldi}.
However, previous flux-tube measurements have only
been made for fundamental representation sources.

This paper presents results of the first direct
lattice measurements of ${\cal A}_j$ for static quarks
in different representations of the gauge group.
This work is aimed in part at establishing
the connection Eq. (\ref{tube}) between the string
tension and flux-tube cross-section. As well,
the results obtained here go beyond the flux-tube model,
providing important new information about the
dynamics underlying confinement.

For example, if confinement is due to a bulk property
of the QCD vacuum, such as a vacuum pressure,
then ${\cal A}_j$ is expected to increase with representation
(a natural ``mechanical'' response of a ``medium'' to the
injection of more intense fields). Consequently,
in such a scenario, the string tension $\sigma_j$ is expected
to increase with representation less rapidly than
the quark Casimir.
This situation is realized in a wide class of
phenomenological models, including
bag models \cite{HasKuti,MITBag}, models based on a
description of the QCD vacuum as a
color-dielectric medium \cite{Dielectric},
and some models of confinement based on vacuum
condensation \cite{Scalar}.

The MIT bag \cite{MITBag} is typical of this
general class of models. In the case of
the heavy quark-antiquark ($Q_j \overline Q_j$) system
described above, the magnitude of the color-electric field
is determined by a balance between the pressure generated
by the field and an external ``bag'' pressure $B$ \cite{HasKuti},
$\case{1}/{2} E_j^2 = B$. It follows that
${\cal A}_j \propto Q_j$, and the
resulting string tension $\sigma_j$ also scales
as the square root of the Casimir,
$\sigma_j \propto (j(j+1))^{1/2}$.

In fact, the phenomenological models discussed above are not
compatible with lattice simulations, which have shown that the
string tension actually scales to an excellent approximation
like the Casimir of the representation. This has been
observed for both SU(2) and SU(3) gauge groups \cite{LGT4D}.
In the context of the flux-tube model, this suggests
that the cross-section ${\cal A}_j$ is {\it independent\/}
of representation.

Direct measurements of the flux-tube cross-section
for heavy quark-antiquark ($Q_j \overline Q_j$) sources
are obtained here in three representations of three-dimensional
SU(2) lattice gauge theory:
fundamental ($j=1/2)$, adjoint ($j=1$), and quartet ($j=3/2$).
A more thorough check of scaling and finite volume effects
is achieved by working in three dimensions than would be
obtained (with the same computing power) in four dimensions.
We think that our results are relevant to the problem
of confinement in four-dimensional QCD. In particular,
previous lattice studies have shown that the string tension
scales like the Casimir of the representation
in three dimensions \cite{Amb3D,Mawh},
as well as in four dimensions \cite{LGT4D}.
Moreover, the flux-tube picture of
confinement is qualitatively the same in both
three and four dimensions. It is therefore reasonable to expect
that the qualitative features of flux-tubes reported here
would be reproduced in four dimensions. Of course, such a
calculation can and should be done.

We find ${\cal A}_j \approx {\rm constant}$ for
the three representations, to within about 10\%
(a rough estimate of the overall quality of our data).
This is consistent with the flux-tube picture, given that
the string tension scales like the Casimir of the representation
(which is confirmed here up to the quartet representation).
Several additional qualitative features of the flux-tube picture
are also verified.

These results suggest a connection between confinement in QCD
and the physics of a dual superconductor.
Indeed, if a multiply-charged monople would be
inserted into an ordinary (type II) superconductor, all the
quanta of magnetic flux would be carried by a single flux-tube,
whose diameter is fixed by the penetration
depth \cite{Nielsen}. A pair of monopoles of opposite sign
would therefore be confined, with a string tension that
would scale like the squared-charge.

It is well-known that dual superconductivity
(magnetic monopole condensation) results in
confinement of electric charges in compact QED in
three-dimensions (QED$_3$) \cite{Polyakov}.
A simple extension of the analytical calculation
of Ref. \cite{Villain} in the Villain approximation
to the Wilson action, to include Wilson loops for
multiply-charged sources, demonstrates that the
string tension scales like the squared-charge.

We have performed lattice simulations of singly- and
doubly-charged Wilson loops in compact QED$_3$, and our
results confirm the expected scaling properties
of the string tension and flux-tube cross-section.
The potential is found to scale like the squared-charge
to within a few percent, and the
flux-tubes in the two cases have the same cross-section
to within about 10\%. The results of our
three-dimensional SU(2) and U(1) simulations
taken together lend some support, albeit indirectly,
to the dual superconductor picture of
confinement in four-dimensional QCD \cite{Dual}.

\section{\bf Method}

To begin with, we consider the three-dimensional SU(2)
lattice theory. Wilson loops are used to introduce
static $Q_j \overline Q_j$ sources. Lattice measurements of the
color-electric and -magnetic fields generated by these
sources are obtained from correlators ${\cal F}^{\mu\nu}_j$
of plaquettes with a Wilson loop
\begin{equation}
   {\cal F}^{\mu\nu}_j(x) \equiv - {\beta \over a^3}
   \left[
   { \left\langle W_j
     \case{1}/{2} {\rm Tr} \, U^{\mu\nu}(x) \right\rangle
     \over \left\langle W_j \right\rangle }
   - \left\langle \case{1}/{2} {\rm Tr} \, U^{\mu\nu} \right\rangle
  \right] ,
\label{Fexact}
\end{equation}
where $U^{\mu\nu}(x)$ is the plaquette located at $x$
(measured relative to the center of the Wilson loop),
and $W_j$ is the normalized trace of the Wilson loop in
the $j$-th representation:
\begin{equation}
   W_j \equiv {1 \over 2j+1} {\rm Tr}
   \left\{ \prod_{l \in L} {\cal D}_j[ U_l ] \right\} .
\label{Wj}
\end{equation}
${\cal D}_j[U_l]$ denotes an appropriate irreducible
representation of the link $U_l$, and $L$ the closed loop.
$\beta \equiv 4 / (g^2 a)$, where the coupling constant $g$
has dimensions of $({\rm mass})^{1/2}$ in three dimensions.

In the continuum limit, the trace of a $1\times1$ plaquette
is by construction independent of representation
(up to overall normalizations). As in several previous lattice
calculations of higher representation Wilson loops
(cf. Refs. \cite{LGT4D,Amb3D,Mawh}),
we use the action expressed in terms of links in
the fundamental representation to perform simulations
at arbitrary $\beta$. The trace of the
plaquette $U^{\mu\nu}$ in the fundamental representation
is also used to compute the correlators of Eq. (\ref{Fexact}).

In the continuum limit the correlator ${\cal F}^{\mu\nu}_j$
corresponds to the expectation value of the square of
the Euclidean field strength
$F^{a\mu\nu} = \partial^\mu A^{a\nu} - \partial^\nu A^{a\mu}
+g \epsilon^{abc} A^{b\mu} A^{c\nu}$,
\begin{equation}
   \lim_{\beta\to\infty} \, {\cal F}^{\mu\nu}_j =
   \case{1}/{2}
   \Bigl\langle \sum_a ( F^{a\mu\nu} )^2
   \Bigr\rangle_{Q_j \overline Q_j}
 - \ \ \case{1}/{2}
   \Bigl\langle \sum_a ( F^{a\mu\nu} )^2 \Bigr\rangle_0
\label{Fa0}
\end{equation}
where the expectation value
$\left\langle \ldots \right\rangle_{Q_j \overline Q_j}$
is taken in a state with external sources in the
$j$-th representation, and
$\left\langle \ldots \right\rangle_0$
is the vacuum expectation value.

To compute the energy density, the Euclidean 3-axis is identified
with a temporal side of the Wilson loop, and the 1-axis with
a radial side. We separate contributions to the
total energy density ${\cal E}_{\rm tot}$
corresponding to the two spatial components of the
color-electric field (in the directions parallel
and perpendicular to the line joining the quarks), and
the color-magnetic field (a scalar in three dimensions):
\begin{equation}
   {\cal E}^{\rm tot}_j(x) = {\cal E}^{\parallel}_j(x)
   +  {\cal E}^{\perp}_j(x)  +  {\cal E}^B_j(x),
\label{Etot}
\end{equation}
where
\begin{eqnarray}
   & & {\cal E}^{\parallel}_j(x) \equiv - {\cal F}^{13}_j(x) ,
\nonumber \\
   & & {\cal E}^{\perp}_j(x) \equiv - {\cal F}^{23}_j(x) ,
\\
   & & {\cal E}^B_j(x) \equiv {\cal F}^{12}_j(x) .
\nonumber
\label{Ecmpts}
\end{eqnarray}
Notice the relative minus sign between the electric and magnetic
components of the Euclidean energy density.

Previous flux measurements for fundamental representation
sources have been made in four dimensions
\cite{Flux1,Flux2,Sommer,HayWos,HayWosFt,Cooling},
and in three dimensions \cite{Caldi}.
Following Haymaker and Wosiek \cite{HayWos}, we achieve a
significant enhancement in the signal to noise for the
correlators by replacing Eq. (\ref{Fexact}) with:
\begin{equation}
   {\cal F}^{\mu\nu}_j(x) \approx - {\beta \over a^3}
   \left[
   { \left\langle W_j
     \case{1}/{2} {\rm Tr}
     \left\{ U^{\mu\nu}(x) - U^{\mu\nu}(x_R) \right\}
     \right\rangle
     \over \left\langle W_j \right\rangle }
  \right] ,
\label{Fapprox}
\end{equation}
where $x_R$ is a reference point chosen far enough from
the Wilson loop that the factorization
$\left\langle W_j U(x_R) \right\rangle \simeq
 \left\langle W_j \right\rangle \left\langle U \right\rangle$
is satisfied.
As in Ref. \cite{HayWos}, we find that this happens well within the
lattice volume. We verified explicitly that
the right-hand side of Eq. (\ref{Fapprox}) is insensitive,
within our statistical errors, to variations in $x_R$
over a wide range (when measurements are made for $x$ in a region
around the Wilson loop of sufficient size to suit our purposes).
The results presented here were obtained
with $x_R$ taken at a distance of half the lattice size
from the center of the Wilson loop, in the direction
transverse to the plane of the loop. The advantage to using
Eq. (\ref{Fapprox}) is that the fluctuations in
the product $W_j U^{\mu\nu}$, due mainly to the Wilson loop,
tend to cancel in the vacuum subtraction when
computed configuration by configuration.

Another reduction in the statistical errors is
readily achieved by performing some link integrations
analytically, following the multihit procedure introduced
by Parisi, Petronzio and Rapuano \cite{Parisi}.
Consider a link variable $U_l$ which appears linearly
in the observable of interest.
The simplest analytical integration over $U_l$
takes account of nearest neighbor couplings in the action:
\begin{equation}
   \int [d U_l] {\cal D}_j[U_l]
   e^{ \beta {\rm Tr} ( U_l K_l^\dagger ) / 2 }
   = { I_{2j+1}(\beta k_l) \over I_1(\beta k_l) }
   {\cal D}_j[V_l]
   \int [d U_l]
   e^{ \beta {\rm Tr} ( U_l K_l^\dagger ) / 2 }
\label{var1}
\end{equation}
where $K_l$ is the sum of the four ``staples'' coupling to
the link of interest $U_l$, and
\begin{equation}
   k_l V_l \equiv K_l, \quad
   {\rm det} \, V_l = 1 .
\label{kl}
\end{equation}

A further variance reduction has been developed by
Mawhinney \cite{Mawh}, which takes account of effective
next-to-nearest neighbor interactions with the
link of interest%%%
\footnote{Mawhinney derived next-to-nearest neighbor
variance reductions for fundamental and adjoint representations
by employing an axial gauge-fixing \cite{Mawh}.
We have generalized his result to arbitrary
representations without gauge fixing.}
\begin{equation}
   \int [d U_l] {\cal D}_j[U_l]
   e^{ - \beta S }
   = { I_{2j+1}(\beta k'_l) \over I_1(\beta k'_l) }
     { I_{2j+1}(\beta k_l) \over I_1(\beta k_l) }
   {\cal D}_j[V'_l V_l]
   \int [d U_l] e^{ - \beta S } ,
\label{var2}
\end{equation}
where $S$ is the action [only next-to-nearest neighbor
couplings to $U_l$ are relevant in Eq. (\ref{var2})], and
\begin{equation}
   k'_l V'_l  \equiv  \sum_{\hat l_\perp}
   U_{\hat l_\perp}(x)
   \sum_{\hat \mu \neq \hat l_\perp}
   U_{l \hat\mu}(x + \hat l_\perp)
   \, U_{\hat l_\perp}^\dagger(x) ,
   \quad {\rm det} \, V'_l \equiv 1 .
\label{kpl}
\end{equation}
$\hat l_\perp$ are four unit vectors perpendicular
to $\hat l$. The oriented plaquette
$U_{l \hat\mu}(x + \hat l_\perp)$ is computed
with the link $U_l(x + \hat l_\perp$) appearing
first on the left [$x$ is the position of the base
of the link $U_l$ in Eq. (\ref{var2})].
The sum over unit vectors $\hat l_\perp$ and $\hat\mu$
in Eq. (\ref{kpl}) is taken over both parallel and antiparallel
orientations with respect to a set of fixed basis vectors.
An integration over the four links $U_l(x + \hat l_\perp)$
is implicit in Eq. (\ref{var2}).

The second order variance reduction of Eq. (\ref{var2}) cannot
be applied to the links in the corners of the Wilson loop,
since some links would then appear more than once in the integrand.
Likewise, the first-order variance reduction Eq. (\ref{var1})
can only be applied to one link in a corner. Further
restrictions apply to calculations of the plaquette correlators.

For Wilson loops with less than six links on a side, we
use the first-order variance reduction Eq. (\ref{var1})
for all links in the loop, except for one link at each corner,
where no variance reduction is used. A plaquette correlator
can be measured simultaneously provided that
all sides of the plaquette are at least one node
from the sides of the Wilson loop.

For Wilson loops of size $6 \times 6$ or larger, we minimize
the variance by using a combination of first- and second-order
variance reductions. Equation (\ref{var1}) is appplied
to the first link and to the second-to-last link on each
(oriented) side of the Wilson loop; Eq. (\ref{var2})
is applied to all other links, except the last link on
each side, where no variance reduction is used.
In this case, a correlator can be measured simultaneously
only if all sides of the plaquette are at least two nodes from
the sides of the Wilson loop.
[This variance reduction scheme can also be applied to Wilson loops
as small as $4 \times 4$ if measurements of correlators
near the center of the loop are not desired. This scheme
is significantly more effective than the one employed in
Ref. \cite{Mawh}, which uses only second-order variance reductions.]

For large $\beta$, $k_l \sim 4$ and $k'_l \sim 12$;
Eq. (\ref{var2}) then provides an estimate of the reduction
$v_{\rm red}$ in the variance of a Wilson loop
using the above scheme, compared to the variance when
only ``unreduced'' links are used
(cf. Refs. \cite{MichaelVar,Mawh})
\begin{equation}
   v_{\rm red} \approx
   \left( { I_{2j+1} (12 \beta) \over I_1 (12 \beta) }
  \right)^{ (2T + 2R - 12) }
   \left( { I_{2j+1} (4 \beta) \over I_1 (4 \beta) }
  \right)^{ (2T + 2R - 4) } .
\label{vred}
\end{equation}
For example, the variance reduction for a Wilson loop
of size $6\times6$ in the quartet representation
at $\beta=10$ is estimated to be a factor of $\approx 90$.
Our numerical results are consistent with Eq. (\ref{vred}).

The trace of an element of the group in the $j$-th
representation can be expressed in terms of its
trace in the fundamental representation using trigonometric
relations among the group characters. In the case of the
adjoint and quartet representations \cite{Redlich}:
\begin{eqnarray}
    & & W_{3/2} = 2 W_{1/2}^3 - W_{1/2} , \nonumber \\
    & & W_1 = \left( 4 W_{1/2}^2 - 1 \right) / 3 .
\label{Trace}
\end{eqnarray}
Hence one need only compute the Wilson loop in the
fundamental representation, using the ``unreduced'' links $U_l$,
or the ``reduced'' elements $V_l$, $V'_l$ of Eqs. (\ref{var1})
and (\ref{var2}), as the case may be.
The Wilson loops in higher representations then follow
from Eq. (\ref{Trace}).
The Bessel functions for the analytical integrations
are tabulated separately for the three representations.

\section{\bf Results and Discussion}

Our main results were obtained on a $32^3$ lattice at
$\beta =10$ (which is well within the scaling region
for the string tension on a lattice of this size \cite{Mawh}).
Wilson loops and plaquette correlators
were calculated in the three representations $j=1/2$, $1$, and $3/2$
for all loops of sizes $T\times R$ from
$3 \times 4$ to $8 \times 8$ (these observables were
measured in groups in several separate runs).
Some additional data was taken at $\beta = 14$ in order
to check for scaling of the physical flux-tube dimensions.
A standard heat-bath algorithm was employed.
More than 10,000 sweeps were typically used for thermalization.
2,000 measurements were made, taking 20 sweeps between measurements.
The resulting integrated autocorrelation times $\tau_{\rm int}$
for the Wilson loops generally satisfy $\tau_{\rm int} \alt 1$,
consistent with the results of a systematic study made in
Ref. \cite{Mawh}. Estimates of the statistical errors were obtained
using the jackknife method. However, measurements of different
observables (and of a given observable in the three
representations) tend to be strongly correlated,
since many Wilson loops and plaquette correlators
were measured simultaneously on a given lattice.

The quartet representation is much more
difficult to measure than the two lower representations,
due to the exponential suppression of the Wilson loop with
the $Q_j \overline Q_j$ potential, which is found to scale
with the Casimir of the representation.
Energy density measurements in the quartet case
obtained from loops larger than about
$6\times6$ are of poor quality, although these
data are consistent with conclusions drawn from results
obtained from smaller loops.

Representative data for Wilson loops in the three representations
are shown in Fig. \ref{FigWloops}. Earlier
studies have shown that the potentials
scale with the Casimir of the representation at
essentially all lengths scales $R$ \cite{Amb3D,Mawh}.
This is made evident in Fig. \ref{FigWloops}, where
the logarithms of the Wilson loops are scaled
by a ratio of Casimirs,
\begin{equation}
    c_j \equiv {3 / 4 \over j(j+1)} .
\label{cj}
\end{equation}
The quantity $-\ln \langle W_j(T,R) \rangle / T$, which
extrapolates to the $Q_j \overline Q_j$ potential $V_j(R)$
in the limit $T\to\infty$, is found to scale as $j(j+1)$
to within a few tenths of a percent at all $T$ and $R$
considered here. A simple extrapolation
of the data using
$V_j(R) \approx \ln[ \langle W_j(T_{\rm max}-1,R) \rangle
/ \langle W_j(T_{\rm max},R) \rangle ]$,
where $T_{\rm max}$ is the largest $T$ value
in the data set, gives agreement to a few tenths of a percent
with the results of a careful statistical analysis of
fundamental and adjoint Wilson loops reported in Ref. \cite{Mawh}.

Several attributes of the plaquette correlators were measured.
To begin with, results for the fundamental and adjoint
representations are presented.
The correlators were measured over a range of distances
$x_\perp$ from the center of the Wilson loop,
in the direction normal to the plane of the loop.
Results for the $T\times R = 8\times6$ loop
are shown in Fig. \ref{FigExp}.
The cross-sections of the fundamental
and adjoint representation flux-tubes are indistinguishable
within statistical errors. This is true for all
Wilson loops that were considered.
For example, the $T$ evolution of ${\cal E}^\parallel_j$
for $R=6$ Wilson loops is illustrated in Fig. \ref{FigET}.
As observed in Refs. \cite{Sommer,HayWosFt}, the
plaquette correlators are more sensitive to higher states than
the Wilson loop. Our data are consistent with a one-excited-state
parameterization given in Ref. \cite{HayWosFt}.

Figure \ref{FigExp} demonstrates that the component of the
color-electric field parallel to the line joining the charges
dominates the energy, as assumed in the flux-tube model.
The magnetic energy turns out to be negative, which has also been
observed in four-dimensional SU(2) lattice theory \cite{HayWos}.
The formation of a well-defined flux-tube is demonstrated
by measurements of ${\cal E}^\parallel_j$ in the plane
of the Wilson loop. Figure \ref{FigExl} shows
${\cal E}^\parallel_j$ for the $T\times R = 6\times 8$ loop
as a function of the longitudinal distance $x_\parallel$
of the plaquette centroid from the center of the loop.
Notice the approximate symmetry of the energy density
about the center of the loop.
The formation of the flux-tube is further illustrated
in Fig. \ref{FigExl0R}, where ${\cal E}^\parallel_j$
is shown as a function of the radial separation
$R$ of the Wilson loop (for fixed $T=6$)

A stringent test of energy density calculations
using Eq. (\ref{Fapprox}) is provided a sum rule
derived by Michael \cite{SumRules}
\begin{equation}
   a^2 \sum_{\scriptstyle \vec x}
   {\cal E}^{\rm tot}_j(\vec x) = V_j(R) .
\label{SumEtot}
\end{equation}
The analogous sum rule in four-dimensional SU(2) was
studied in detail by Haymaker and Woseik \cite{HayWos}.
The flux-tube picture suggests a related sum rule
that is much simpler to measure. If the interaction energy
is dominated by a constant color-electric field along the line
joining the charges (as expected in the limit of quark separations
much greater than the flux-tube thickness), then the integral
of the energy density along one transverse ``slice'' of the
flux-tube should equal the string tension
[cf. $\sigma_j = \lim_{R\to\infty} (V_j(R) - V_j(R-a)) / a$]:
\begin{equation}
   a \sum_{\vert x_\perp \vert}
   {\cal E}^\parallel_j(x_\perp,x_\parallel={\rm fixed})
   \approx \sigma_j ,
\label{SumExp}
\end{equation}
where the sum is taken over positive and negative
distances $x_\perp$ from the plane of the Wilson loop.

Our results are in good agreement with Eq. (\ref{SumExp}).
Figure \ref{FigSum} shows the left-hand-side
of this equation for the $T\times R = 8\times 6$ loop
in the fundamental and adjoint representations,
using a variable cutoff $x_\perp^*$ on the sum.
The right-hand-side of Eq. (\ref{SumExp}) is illustrated
by the dashed line in Fig. \ref{FigSum} (our estimates of
the string tension in the two representations
agree with Ref. \cite{Mawh} to within a few percent).
These results again demonstrate that the flux-tube
cross-sections for the fundamental and adjoint repesentations
are indistinguishable within statistical errors.

Figures \ref{FigExp}--\ref{FigSum} also demonstrate that the
local energy densities scale to a good approximation
like the Casimir of the representation {\rm throughout\/}
the flux-tube. This is a very strong test of the validity
of the flux-tube model for ${\cal A}_j = {\rm constant}$.
Since the magnitude of the color-electric field varies
across the flux-tube cross-section (cf. Fig. \ref{FigExp}),
a proper determination of the numerical value of ${\cal A}_j$
should be made in terms of expectation values of the
color-field, as described below Eq. (\ref{tube})
[some prescription for defining the Abelian projection
of the color-field would also be required].
However, a rough estimate ${\cal A}_j \approx 8 a$
inferred from Fig. \ref{FigSum}
is consistent with Eq. (\ref{tube}),
given the estimate of the string tension
$c_j \sigma_j \approx 0.14 g^4$ (cf. Ref. \cite{Mawh}).

The cross-section is also found to be approximately
independent of $R$. A similar conclusion was reached
in four dimensions in Ref. \cite{HayWos}.
On the other hand, the cross-section
in the strong coupling limit in four dimensions is found
to increase logarithmically with $R$ \cite{Luscher}.
Within statistical errors the range in $R$ considered here
is not sufficient to rule out such a weak dependence on
the radial separation.

Our measurements of the quartet representation ($j=3/2$)
correlators are consistent with the above results. We compare
data in the three representations
taken from the $T\times R = 5 \times 6$ Wilson loop:
energy density profiles transverse to the
flux-tube are shown in Fig. \ref{FigEQ}, and the sum
rules Eq. (\ref{SumExp}) in Fig. \ref{FigSumQ}.
Data obtained from larger Wilson loops are consistent
with these results although, as mentioned above,
the quartet data for larger loops are of poor quality,
due to an exponential suppression of the Wilson loop
with the Casimir of the representation.

We checked for scaling of the physical flux-tube
dimensions by running at $\beta=14$.
The sum rule Eq. (\ref{SumExp}) for the fundamental
representation is compared at the two values of $\beta$ in
Fig. \ref{FigScaling}. The cutoff $x_\perp^*$ on
the sum is expressed here in units of the physical
coupling constant $g$. The data at $\beta = 10$ are
for a $6\times 6$ Wilson loop, while the data at $\beta =14$
are for an $8\times 8$ loop.
These Wilson loops have roughly the same dimensions in
physical units ($T$ and $R$ measured in units of $1/g^2$).

These results show good evidence for
scaling in the energy density and flux-tube cross-section
(scaling is also observed in our adjoint and quartet
representation data). However, the factorization assumed in
Eq. (\ref{Fapprox}) breaks down in the $\beta=14$
data at the largest values of the cutoff $x_\perp^*$
shown in Fig. \ref{FigScaling} ($x_\perp^* \sim 9.5$
in lattice units, to be compared with
$x_R = 16$); the sum is found to diverge linearly with $x_\perp^*$
at large cutoffs. A similar behavior was observed in
four-dimensional lattice calculations in Ref. \cite{HayWos},
where a correction for this effect was proposed.
Nevertheless, scaling of the cross-section is clearly
supported by data in the region $x_\perp^* \alt 1.8 / g^2$.

As described in the Introduction, the results
of our SU(2) simulation suggest a connection between
confinement in QCD and the physics of a dual
superconductor. In this connection, we have calculated
Wilson loops $W_n$ in compact QED$_3$ for
singly- and doubly-charged sources ($n=1,2$):
\begin{equation}
   W_n \equiv {\rm Re} \, \prod_{l \in L} (U_l)^n ,
\label{WjQED}
\end{equation}
where the phase $U_l$ for the link $l$ defines the
singly-charged representation (i.e., $U_l$ is the
phase used to compute the Wilson action).
The string tension is expected to scale like
the squared-charge, as demonstrated by
an extension of the Villain approximation
used in Ref. \cite{Villain} to include multiply-charged Wilson
loops. As in an ordinary superconductor, the flux-tube
cross-section is expected to be independent of source charge.

Wilson loops and plaquette correlators were measured
on a $32^3$ lattice at $\beta=2.4$. More than
10,000 sweeps were used for thermalization, and
1,000 measurements were made (90 sweeps were taken
between measurements). Variance reduction methods
similar to those used in our SU(2) simulations
were employed.%%
\footnote{The analytical integrals given
in Eqs. (\ref{var1}) and (\ref{var2}) are easily adapted to the
U(1) theory. The SU(2) link ${\cal D}_j[U_l]$ becomes $(U_l)^n$,
the Bessel function ratios $I_{2j+1}(x) / I_1(x)$
are replaced by $I_n(x) / I_0(x)$, and ${\rm det}(V)$ becomes
${\rm abs}(V)$.}

Results for the Wilson loops are given in Fig. \ref{FigQloops}.
Some data for the triply-charged Wilson loop ($n=3$)
are also shown (useful measurements of the plaquette
correlators for $n=3$ would require much larger statistics).
Estimates of the potential for $n=1$ obtained from a simple
extrapolation of these data are consistent with results presented
in graphical form in Ref. \cite{QED3MC}.

We find that $-\ln \, \langle W_n(R,T) \rangle / T$, which
extrapolates to the potential $V_n(R)$ in the limit $T\to\infty$,
scales like $n^2$ to within about 2\% for all $T$ and $R$ that were
considered, in good agreement with the expected
scaling properties of the string tension. However,
the deviation from $n^2$ scaling is about an order
of magnitude larger than the statistical errors in the data.
String vibrational modes are known to make a significant
contribution to $V_1$ in the range of $R$ considered here
(lattice simulations \cite{QED3MC} are in agreement with
theoretical expectations \cite{Luscher}).
Simple arguments \cite{Amb3D} suggest that
the vibrational term in $V_n$ may scale like $n$, which could
account for the small deviation from $n^2$ scaling
in the logarithms of the Wilson loops.

The energy sum rule analogous to Eq. (\ref{SumExp}) for the
$T\times R = 5 \times5$ loop is shown in Fig. \ref{FigQEDSum}.
The dashed line shows the $n=1$ string tension
taken from Ref. \cite{QED3MC}.
These results provide the first direct evidence from
lattice simulations that the flux-tube cross-section
in compact QED$_3$ is independent of the source charge,
the expected behavior in the case of a (dual) superconducting medium.

\section{\bf Summary}
The first direct measurements of the flux-tube cross-section
as a function of representation in SU(2) lattice gauge theory
were made. We found ${\cal A}_j \approx {\rm constant}$,
to within about 10\% (a rough estimate of the overall quality
of our data) for the three representations $j=1/2$, $1$, and $3/2$.
Our results are consistent with a connection between the
string tension and cross-section suggested by a simplified
model of flux-tube formation,
$\sigma_j = g^2 j(j+1) / (2 {\cal A}_j)$,
given that the string tension scales like the
Casimir $j(j+1)$, as observed in previous lattice
studies in both three and four dimensions
(and confirmed here up to the quartet representation).
We also confirmed several additional qualitative
features of the flux-tube picture of color-electric confinement.
These results can be used discriminate among phenomenological
models of the physics underlying confinement. For example,
many models in which confinement is due to a bulk property of
the QCD vacuum (such as a vacuum pressure) predict a
sufficiently rapid increase in ${\cal A}_j$ with representation
as to be incompatible with the results obtained
from our lattice simulations.

We also made flux-tube measurements in compact QED$_3$,
which exhibits electric confinement due to magnetic
monopole condensation. We considered singly- and doubly-charged
Wilson loops. The string tension was found to scale
like the squared-charge, and the flux-tube cross-section
was found to be independent of the charge, to a good
approximation. The results of our three-dimensional
SU(2) and U(1) simulations taken together lend some support,
albeit indirectly, to a conjecture that the
dual superconductor mechanism underlies confinement
in compact gauge theories in both three and four
dimensions. This conclusion is also supported by the results
of a recent study of dual Abrikosov vortices in an Abelian
projection of SU(2) lattice gauge theory in
four dimensions \cite{HayVortex}.
Flux-tube measurements in four-dimensional SU(2)
gauge theory similar to those reported here should
be made in order to further explore this possibility.

\acknowledgments

This work was
supported in part by the Natural Sciences and Engineering
Research Council of Canada.

%%%%%%%%%%%%%%%%%%%%%%%
%%% FIGURE CAPTIONS %%%
%%%%%%%%%%%%%%%%%%%%%%%

\figure{$T$ evolution of Wilson loops in three
representations of SU(2) lattice gauge theory:
$j=1/2$ ($\circ$), $j=1$ ($\Box$), and $j=3/2$ ($\triangle$).
$c_j$ is a ratio of Casimirs, defined in
Eq. (\ref{cj}). The quantity $-\ln\langle W_j(T,R)\rangle / T$
extrapolates to the $Q_j \overline Q_j$ potential
in the limit $T\to\infty$.\label{FigWloops}}

\figure{Energy density profiles transverse to
the plane of the $T\times R =8\times6$ Wilson loop
[$j=1/2$ ($\circ$), and $j=1$ ($\Box$)].
These results are an average over plaquettes
with centroids at distances $\pm x_\perp$
transverse to the plane of the loop.\label{FigExp}}

\figure{$T$ evolution of ${\cal E}^\parallel_j(x_\perp=0,2a)$
for Wilson loops with $R=6$
[$j=1/2$ ($\circ$), $j=1$ ($\Box$)].\label{FigET}}

\figure{Energy density profile in the plane of the
$T\times R = 6\times8$ Wilson loop
[$j=1/2$ ($\circ$), $j=1$ ($\Box$)].
$x_\parallel$ is the distance of the centroid of the plaquette
from the center of the Wilson loop. The radial sides
of the Wilson loop are located at $x_\parallel = \pm 4a$.
Plaquettes with a side touching the Wilson loop cannot be
measured using the variance reduction of Eq. (\ref{var1}),
and are not shown.\label{FigExl}}

\figure{Energy density ${\cal E}^\parallel_j(x_\perp=0,2a)$
as a function of $R$, for fixed $T=6$
[$j=1/2$ ($\circ$), $j=1$ ($\Box$)].\label{FigExl0R}}

\figure{Energy sum rule Eq. (\ref{SumExp}) for
the $T\times R=8 \times6$ Wilson loop
[$j=1/2$ ($\circ$), $j=1$ ($\Box$)].
The sum in Eq. (\ref{SumExp}) is evaluated using
a variable cutoff $x_\perp^*$.
The dashed line shows the scaled string tension
$c_j \sigma_j$, estimated to about 5\%
(cf. Ref. \cite{Mawh}).\label{FigSum}}

\figure{Energy density profile transverse to the plane
of the $T\times R=5\times6$ Wilson loop,
in the fundamental and quartet representations
[$j=1/2$ ($\circ$), $j=3/2$ ($\triangle$)].\label{FigEQ}}

\figure{Energy sum rule Eq. (\ref{SumExp}) for
the $T\times R = 5\times6$ Wilson loop, in all three
representations [$j=1/2$ ($\circ$), $j=1$ ($\Box$),
$j=3/2$ ($\triangle$)]. The sum rule improves with increasing
$T$ and $R$ (cf. Fig. \ref{FigSum}).\label{FigSumQ}}

\figure{Scaling of the energy sum rule Eq. (\ref{SumExp})
for the fundamental representation. The open data points
were taken at $\beta=10$ (for a $6\times6$ Wilson loop),
and the filled points at $\beta=14$ (for an $8\times8$ loop).
The cutoff $x_\perp^*$ is expressed here in units of
the physical coupling constant $g$.\label{FigScaling}}

\figure{$T$ evolution of multiply-charged Wilson loops
in compact QED$_3$ [$n=1$ ($\circ$), $n=2$ ($\Box$),
and $n=3$ ($\triangle$)].\label{FigQloops}}

\figure{Energy sum rule Eq. (\ref{SumExp}) for
singly- and doubly-charged sources in compact QED$_3$,
for the $T\times R = 5\times5$ Wilson loop
[$n=1$ ($\circ$), $n=2$ ($\Box$)].
The dashed line shows the string tension
$\sigma_{n=1}$, estimated to about 10\%
in Ref. \cite{QED3MC}.\label{FigQEDSum}}

\end{document}